\documentclass[conference, twoside]{IEEEtran}
\IEEEoverridecommandlockouts
\usepackage{graphicx}
\usepackage{multirow}
\usepackage{kotex}
\usepackage{cite}
\usepackage{amsmath,amssymb,amsfonts}
\usepackage{algorithmic}
\usepackage{graphicx}
\usepackage{textcomp}
\usepackage{xcolor}
\usepackage{lipsum}
\def\BibTeX{{\rm B\kern-.05em{\sc i\kern-.025em b}\kern-.08em
    T\kern-.1667em\lower.7ex\hbox{E}\kern-.125emX}}
\begin{document}
\sloppy
\newcommand*\concat{\mathbin{\|}}
\newcommand{\cy}[1]{{[\bf \color{red}  {CY: } #1}]}
\newcommand{\js}[1]{{[\bf \color{blue}  {JS: } #1}]}

\title{\vspace{0.5em} Interactive 3D Tangible Display with a High-Speed Stiffness-Variable Jamming Module
% {\footnotesize \textsuperscript{*}Note: Sub-titles are not captured for https://ieeexplore.ieee.org  and
% \author{Chanyoung Ahn\textsuperscript{1}, Jaesung Lee\textsuperscript{1}, and Donghyun Hwang\textsuperscript{\textdagger}
\thanks{
This work was supported in part by Samsung Science and Technology Foundation under Grant SRFC-IT2202-01,  and in part by the Korea Institute of Science and Technology Institutional Program.
(Chanyoung Ahn and Jaesung Lee contributed equally to this work.) (Corresponding author: Donghyun Hwang.)
Chanyoung Ahn, Jaesung Lee, and Donhyun Hwang are with the Center for Robotics Research, KIST, Seoul 02792, South Korea (\{chanyoung.ahn, jay.lee, donghyun\}@kist.re.kr).
% All authors are with Korea Institute of Science and Technology (KIST), Korea ({\tt\footnotesize \{chanyoung.ahn, jay.lee, donghyun\}@kist.re.kr}). {\textsuperscript{\textdagger}}D. Hwang is the corresponding author.
}
}

\author{\IEEEauthorblockN{Chanyoung Ahn}
\IEEEauthorblockA{
\textit{Center for Humanoid Research} \\
\textit{KIST}\\
Seoul, South Korea \\
chanyoung.ahn@kist.re.kr}
\and
\IEEEauthorblockN{Jaesung Lee}
\IEEEauthorblockA{\textit{Center for Humanoid Research} \\
\textit{KIST}\\
Seoul, South Korea \\
jay.lee@kist.re.kr}
\and
\IEEEauthorblockN{Donghyung Hwang}
\IEEEauthorblockA{\textit{Center for Humanoid Research} \\
\textit{KIST}\\
Seoul, South Korea \\
donghyun@kist.re.kr}
}

\maketitle

\begin{abstract}
Multisensory integration, particularly through visual and tactile feedback, plays a crucial role in enhancing audience engagement with artworks. Although recent research has increasingly explored tactile experiences in art, existing systems often lack real-time variable stiffness modulation and depend on bulky mechanical infrastructures. In this work, we propose a novel tangible display based on a magnetic jamming mechanism, enabling real-time, low-noise, and low-voltage stiffness modulation integrated into traditional sculptural artworks. Our system combines visual motion and dynamic tactile feedback within a compact standalone module, allowing audiences to interactively experience variations in the rigidity and form of features such as those found in the traditional Korean mask Hahoetal. This approach offers a new paradigm for interactive art, enabling more immersive, multisensory engagement through the fusion of cultural artifacts and modern technology.
\end{abstract}


\section{Introduction}
\label{sec:intro}
% multisensory interaction in art -> current limitation -> problem 
% Traditional korean art + technology -> new approach
% paper contribution + statement 

 In the fields of arts and media, multisensory integration, such as vision and physical touch, significantly enhances the audience's engagement with art and stimulates their imagination \cite{vi2017not}. Consequently, touch has gained increasing attention in sensory studies, particularly in the context of art \cite{vi2017not, lauwrens2019touch, hayes2017towards}. Despite its recognized importance, there remains a lack of research on tactility as an aesthetic experience \cite{lauwrens2019touch, gallace2011tactile}. Furthermore, while some artworks attempt to integrate sensory argumentation, they only address art forms that lack interactivity with viewers such as non-variable contact feedback from artworks. 

    \begin{figure}[t]
    \centering
    \includegraphics[width=\columnwidth]{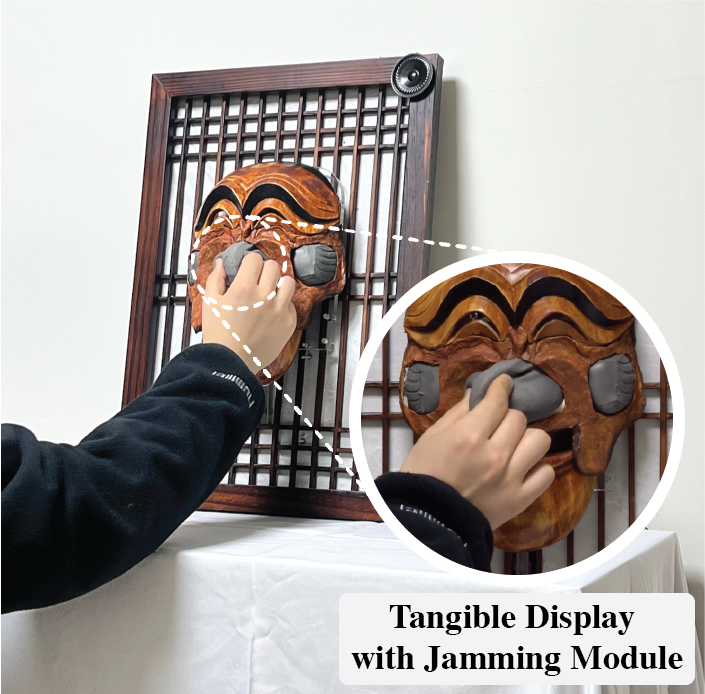}
    \caption{The proposed tangible user display illustrates various levels of rigidity achieved through a magnetic jamming mechanism. A traditional Korean mask, "Hahoetal," is used to demonstrate this concept. The mask features a soft, human-like nose capable of dynamically adjusting its shape and rigidity feedback, enabling users to appreciate its tactile qualities and interact with it in real-time.}
    \label{img:main}
    % \vspace{-5mm}
    \end{figure}

 To address these limitations, researchers in the field of Human-Computer Interaction (HCI) have explored various approaches to enable variable stiffness in artworks, enhancing the tactile experience during viewers' interaction with art \cite{vi2017not, follmer2013inform, stanley2016deformable}. One such approach, inForm \cite{follmer2013inform}, introduces an interactive user interface that utilizes variable stiffness and supports real-time tangible interactions, allowing users to manipulate and experience dynamic physical representations. However, its system relies on a complex setup of 900 mechanical actuators and a computer, which makes it resource-intensive. Additionally, it is limited to creating pixelated, shape-changing surfaces, restricting its ability to represent finer, more nuanced tactile details. Another approach \cite{stanley2016deformable} employs a combination of particle jamming and pneumatics to produce continuous surfaces capable of forming natural shapes, such as organic objects. While this method offers greater flexibility in creating smoother and more lifelike forms, it requires a bulky air chamber, which generates significant noise and results in slow stiffness adjustment times, limiting its practicality for real-time interaction. 

 % ?
 Despite these efforts, existing research still faces three significant challenges: first, dependence on large mechanical systems; second, difficulty in achieving real-time variable stiffness; and third, difficulty in achieving stable shape fixation after freeform deformation. These limitations restrict the diversity of tactile experiences, particularly in capturing nuanced sensations such as the delicate softness of a human face in sculptural art. Unlike pneumatic jamming systems that require bulky external infrastructure or mechanical actuator arrays that entail high mechanical complexity and cost, our approach leverages a magnetic jamming-based mechanism to achieve rapid stiffness modulation with a compact, low-noise, and low-voltage system. By employing a modular, electrically-driven configuration, our system seamlessly integrates into traditional artworks, preserving both their aesthetic beauty and cultural significance. Furthermore, it supports real-time multisensory interaction—combining tactile feedback through stiffness modulation and visual feedback via mask motion—providing a richer and more dynamic audience experience.

 In this work, we propose a novel tangible display based on a magnetic jamming-based 3D display. Our stand-alone jamming module enables interaction with portraits through integrated visual and tactile feedback, offering an enriched aesthetic experience. This system allows the audience to experience real-time variations in rigidity and to actively modify the shape of the artwork, thereby enhancing immersive engagement (see Figure \ref{img:main}). Unlike previous systems, our portable system operates with a voltage input of less than 25 V and avoids the need for bulky mechanical components.
\section{Related Work}
\label{sec:related}

\subsection{Tangible Displays for Art Interaction}

 In the field of Human-Computer Interaction (HCI), Tangible User Interfaces (TUIs) have emerged as pivotal tools that bridge the physical and digital worlds, enabling intuitive and inclusive interactions with artistic content. By allowing users to engage with digital information through physical manipulation, TUIs leverage our natural abilities to interact with the environment, reducing cognitive load and fostering more immersive art experiences that extend beyond visual perception to include touch and movement \cite{klemmer2006bodies}.
 
Traditional art exhibitions often rely heavily on visual elements, posing challenges for individuals with visual impairments. To address this, institutions have implemented tactile models and multi-sensory exhibits. For instance, the National Museum of Wildlife Art introduced a 3D tactile display of the artwork "Chief \cite{TactileChief2024}" allowing blind and low-vision visitors to engage with the piece through touch and sound. However, these models are static and lack dynamic responsiveness.

Advancements in TUIs have led to the development of dynamic shape displays, which can change form in response to digital information. MIT's Tangible Media Group developed inform \cite{follmer2013inform}, a system consisting of approximately 1,000 pins that can move individually to create topographical renderings of objects. While innovative, such systems often involve high costs, mechanical complexity, and lack variable stiffness capabilities.

In this work, we propose a compact, low-voltage, and low-noise variable stiffness module that can be seamlessly integrated into traditional sculptural artworks such as the Hahoetal mask. Our system enables real-time stiffness modulation through a single electrical input without the need for bulky external devices (see Fig.\ref{img:main}), thus preserving the aesthetic integrity of the artwork. Through this approach, we explore new possibilities for transforming traditional art into interactive, physically dynamic experiences.

\begin{figure*}[t]
    \centering
    \includegraphics[width=0.9\textwidth]{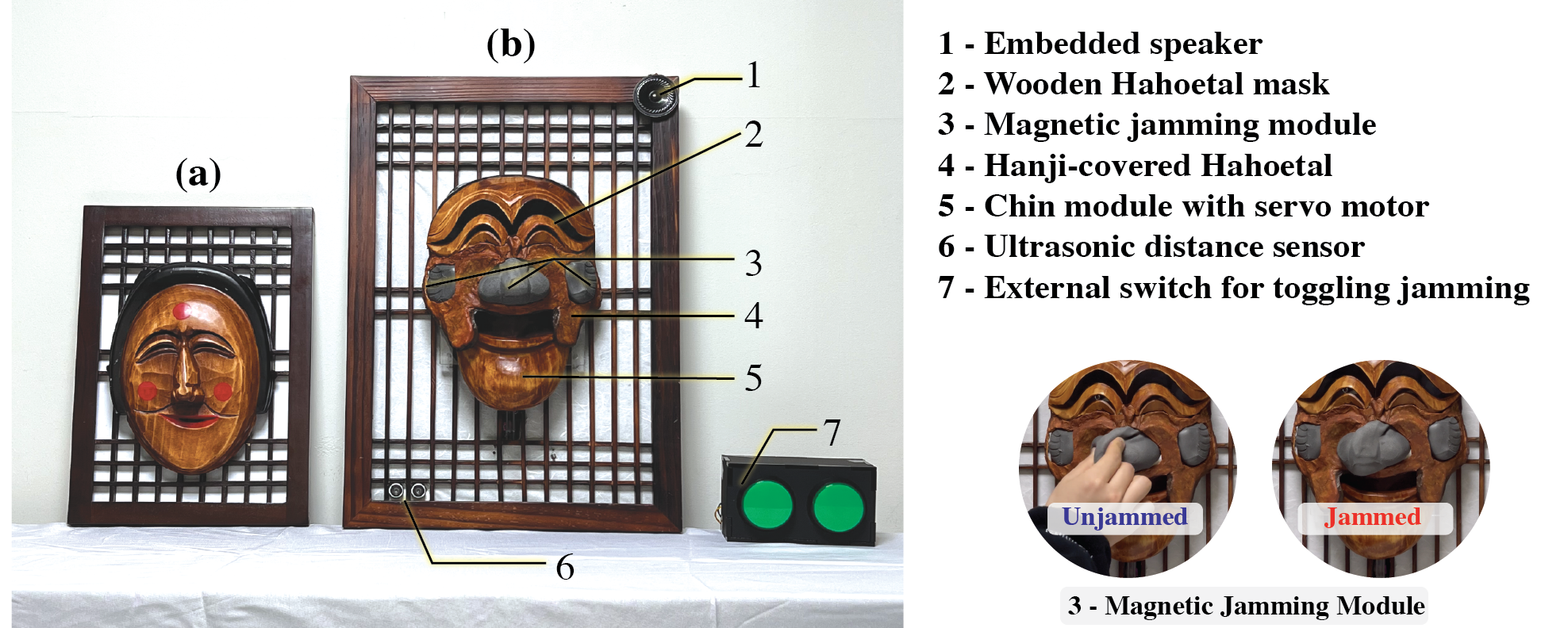}
    \caption{Overview of the proposed interactive artwork based on the Korean traditional \textit{Hahoetal} masks. The left sculpture (a) features a static \textit{Bune} (young woman role) mask that does not respond or deform in real time. The central piece (b) is an interactive \textit{Yangban} (aristocrats) mask that detects viewer presence, initiates speech, and dynamically modulates the stiffness and shape of its nose and cheeks. On the right side, a manual switch allows viewers to activate or deactivate the jamming modules, enabling real-time control over the stiffness transition of the tactile surfaces.}
    \label{fig:overall-structure}
\end{figure*}

\subsection{Jamming Methods}
Granular jamming techniques have gained widespread attention for enabling variable stiffness in soft robotics and adaptive hardware due to their superior shape conformity to irregular surfaces \cite{aktacs2021modeling, yang2021reprogrammable}. Most traditional jamming systems employ pneumatic actuation via vacuum pumps to compress granular materials \cite{amend2012positive, stanley2013haptic}. However, pneumatic systems typically exhibit slow transition time. Sometimes up to tens of seconds when using compact or portable pumps. Additionally, these systems tend to be bulky because of their reliance on external pneumatic components and produce considerable noise and vibrations, which limit their applicability in quieter or miniaturized environments.

To address these drawbacks, other actuation strategies have been explored. Mechanical methods using cables \cite{wockenfuss2022design, choi2021tendon} or clamps \cite{zhou2019discrete} provide alternative jamming actuation but suffer from limited geometric flexibility and relatively slow response times. Electrostatic layer jamming achieves rapid stiffness modulation \cite{wang2019electrostatic} but requires high voltage inputs (exceeding 1000 V) and involves complex fabrication, which hinders wide adoption.

Magnetic field-driven jamming approaches present promising advantages including fast responses, quiet operation, and straightforward electrical control. Nonetheless, prior studies have mainly focused on the underlying magnetic particle physics\cite{valverde2011jamming} or bistable memory \cite{liu2022magnetic} applications rather than systematically developing variable stiffness structures suitable for robotics. Early magnetic jamming grippers exist \cite{varner2023utilization} but lack comprehensive design optimization and detailed performance characterization. Compared to magnetorheological (MR) fluids, magnetic granular jamming generally yields higher absolute stiffness due to the use of larger magnetic particles, though issues such as particle settling and slippage at the interface have posed challenges to effective stiffness modulation.

The proposed magnetic jamming module, based on the design introduced in \cite{song2025stiffening}, combines mixed-size iron particles, a magnetic elastomer membrane, and an electromagnet to enable rapid and precise electrical control of stiffness. This modular design facilitates fast stiffness transitions within approximately 0.1 seconds (see Fig.\ref{img:graph}) with minimal noise. Its repeatable and linear stiffness changes allow for reliable physical interactions in robotic applications. We present a variable-stiffness tangible display as an advanced artistic innovation in robotics.

\subsection{Cultural Technology and Art-Technology Fusion} In the fields of arts and media, sensory engagement—particularly through vision and physical touch—has been recognized as a critical factor in enhancing audience immersion and stimulating imagination \cite{vi2017not, lauwrens2019touch, hayes2017towards}. In this context, recent research has increasingly explored cultural technology, an approach that integrates traditional art forms with modern interactive technologies to expand sensory experiences.

Various studies have employed sensor-based interaction, digital media, and augmented reality (AR) technologies to facilitate new modes of engagement between audiences and artworks. For instance, some exhibitions dynamically alter visual and auditory outputs in response to viewer movements or gestures, while others overlay digital information onto traditional artworks using AR. These efforts have contributed to shifting the role of audiences from passive observers to active participants. However, most of the existing research remains limited to visual and auditory modalities, with relatively little emphasis on tactile interaction with sculptural artworks \cite{lauwrens2019touch, gallace2011tactile}. Moreover, prior systems often rely on bulky mechanical infrastructures or suffer from slow response times, limiting their effectiveness in delivering immediate sensory feedback during user interaction \cite{follmer2013inform, stanley2016deformable}.

To address these limitations, we propose a new form of immersive cultural experience that combines a traditional Korean sculptural artifact, the \textit{Hahoetal} mask, with a multi-sensory interaction system encompassing tactile, visual, and auditory modalities. By enabling users to touch, deform, and interact with the artwork in real time, our approach not only reinterprets traditional art but also explores the potential for cultural heritage to actively engage with audiences through contemporary interactive technologies.

\section{System Overview}
\label{sec:related}

\subsection{Overall architecture}
% \begin{figure*}[t]
% \center
% \includegraphics[width=0.8\textwidth]{Fig/fig2_1.png}
% \caption{

 Our system enables real-time tangible interaction with artworks by dynamically modulating surface rigidity through a magnetic granular jamming (MGJ) mechanism. As illustrated in Fig. \ref{fig:overall-structure}, the system consists of three main components: (1) a magnetic jamming module based on an electromagnet and soft magnetic particles (see Fig. \ref{fig:overall-structure} (b)-3), (2) a control unit that integrates sound-based user interaction capabilities (see Fig. \ref{fig:overall-structure} (b)-1 and 5), and (3) a tangible artwork display embodied by a traditional Korean mask (\textit{Hahoetal}) embedded with the jamming module. When a user approaches the mask, the system detects their presence and initiates interactive behavior, including moving the mouth to invite the user to touch the nose. Upon contact, the nose region initially exhibits low stiffness, allowing users to deform the surface. After a predefined time window, the MGJ mechanism is activated, rapidly increasing the stiffness and preserving the deformed shape. This real-time integration of auditory, visual, and tactile feedback enhances user immersion, enabling a deeper multisensory engagement with the artwork. Furthermore, due to the lightweight and modular design of the jamming module, the artwork remains highly portable, comparable to the handling ease of a conventional picture frame.

\subsection{Hardware Component}

% \begin{figure}
%     \centering
%     \includegraphics[width=\columnwidth]{Fig/fig8_1.png}
%     \caption{Artwork structure}
%     \label{fig:tal-structure}
% \end{figure}

The hardware system integrates three main physical components: the magnetic jamming module, the control electronics, and the display structure of the tangible artwork. The magnetic jamming module comprises a flat-faced electromagnet with a hollow iron core, soft magnetic particles ($> 99.9$\% purity iron), and a deformable magnetorheological (MR) elastomer membrane that encapsulates the particle chamber. The MR membrane serves a dual purpose: to contain particles while adapting its surface shape in response to external forces, enabling high-fidelity deformation and tactile interaction. An internal support structure ensures mechanical stability and efficient force transmission between the deformable surface and the frame. The design emphasizes compactness and modularity, allowing the module to be embedded in various artistic surfaces without compromising the aesthetic or portability of the artwork. The detailed design of the system is presented in Section IV.A.

The control electronics include a microcontroller (e.g., Arduino Mega 2560), a multi-voltage power regulation system (providing 24V, 9V, and 5V DC output), and a driver circuit for dynamic voltage modulation.  The microcontroller monitors external sensors to detect user proximity and interaction events, adjusting the electromagnetic input in real-time to control the surface stiffness dynamically. The control and power cables are routed internally within the frame to preserve the visual appearance and ensure minimal physical interference during the interaction.

The tangible artwork display is based on a lightweight, rigid frame modeled after the traditional Korean mask, \textit{Hahoetal}, with specially designed cavities to accommodate the jamming modules in critical facial regions (nose and cheeks). The mask surface is finished using \textit{Hanji}, a traditional Korean papercraft technique, preserving cultural aesthetics while providing a tactilely rich surface. In addition, the frame incorporates a traditional Korean grid pattern design, which enhances structural rigidity while minimizing overall weight. These integrated hardware components form a cohesive and portable system capable of providing real-time visual and tactile feedback, significantly enhancing audience immersion in interactive art experiences.

\subsection{Control System}
 The control system orchestrates the interaction between the user and the tangible display by integrating sensor input, electromagnetic control, and motor actuation. A microcontroller receives real-time input from an ultrasonic distance sensor, detecting the approach of a viewer toward the artwork. Upon detection, the system initiates a pre-programmed sequence, including audio playback via an embedded speaker and mouth motion driven by a servo motor to invite the viewer to engage physically with the mask. The stiffness of the nose and cheek regions is modulated through an MGJ mechanism. When the user or operator activates an external switch, the microcontroller interprets the input and deactivates the electromagnets by cutting the 24V supply, lowering the surface rigidity and enabling the viewer to deform the mask’s surface. After a predefined time window, the microcontroller reactivates the electromagnets by resupplying 24V, causing the MGJ modules to rapidly transition into a high-stiffness state and preserve the deformed shape.

To support these multimodal interactions, the system employs multi-voltage supplies. A 24V DC output powers the magnetic jamming modules, 9V operates the servo motors responsible for mouth movements, and 5V supplies the ultrasonic distance sensor, external switch, motor controller, and speaker.  All control signals and power lines are routed internally within the frame structure to preserve the external aesthetic integrity of the artwork and ensure robust, unobtrusive interaction. This integrated control architecture enables intuitive, real-time user engagement by coordinating tactile stiffness modulation, visual motion, and auditory feedback through a centralized microcontroller-based system.

\section{Magnetic Jamming Module}
% input data + magnetic field + stiffness variation 
% specification
% Input control (voltage regulation, magnetic field generation)
% Stiffness modulation logic
\subsection{Design and Fabrication}
\begin{figure}
    \centering
    \includegraphics[width=\columnwidth]{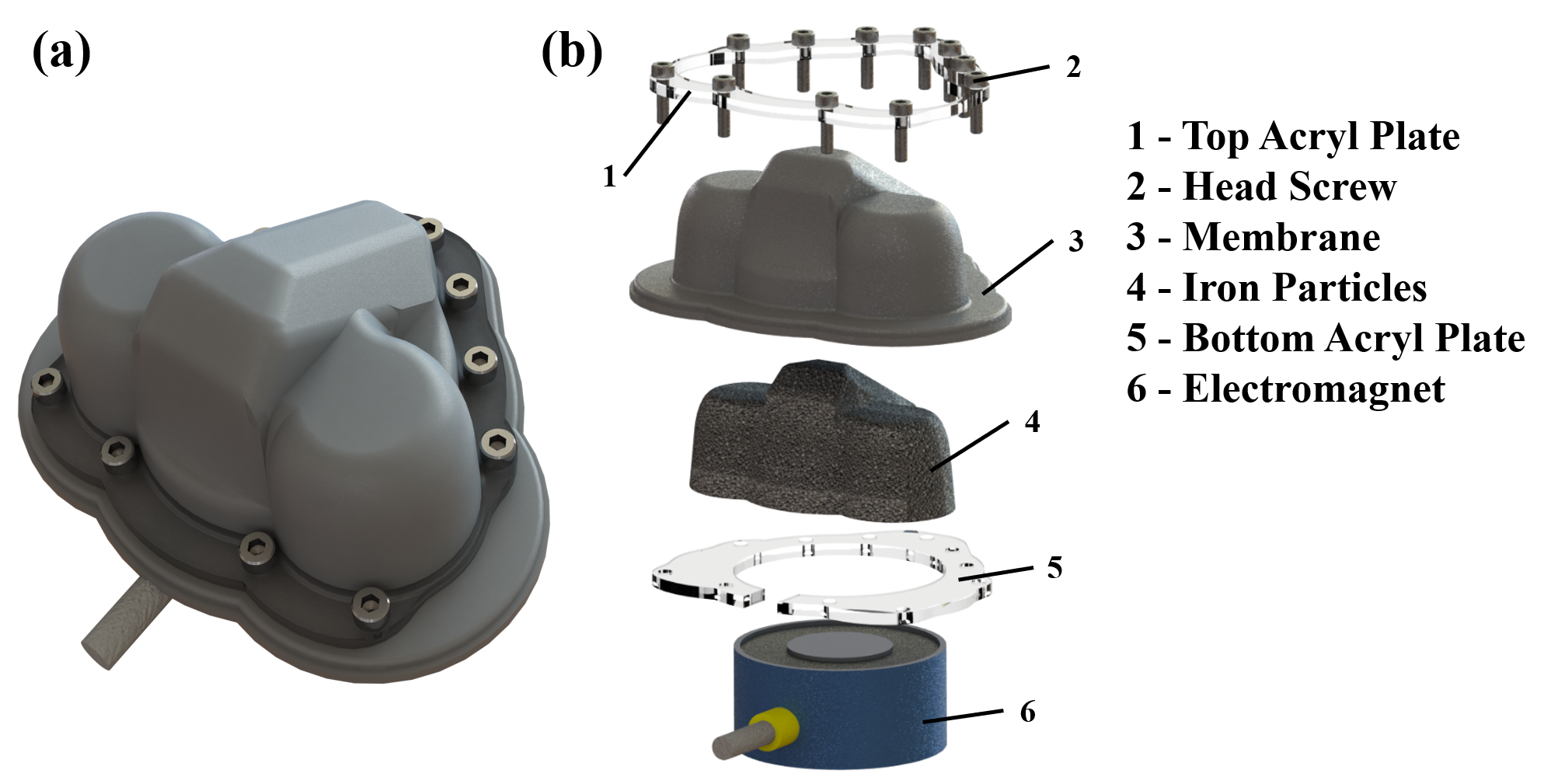}
    \caption{(a) Overall design of the magnetic jamming module. (b) Exploded view and component description of each part.}
    \label{fig:enter-label}
\end{figure}

\begin{figure}
    \centering
    \includegraphics[width=0.8\columnwidth]{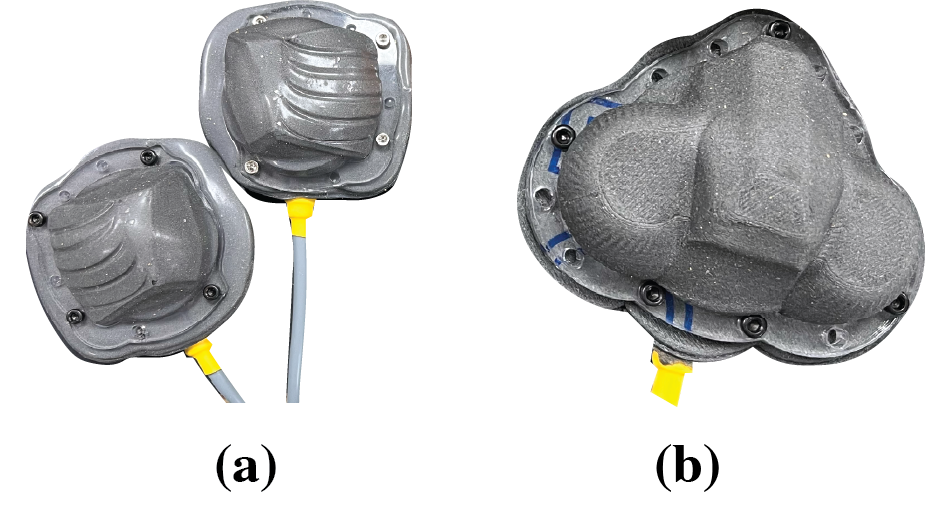}
    \caption{Two customized types of magnetic jamming modules. (a) The cheek module emphasizes strong shape fixation under magnetic actuation. (b) The nose module exhibits a high stiffness variation ratio and enhanced elastic recovery.}
    \label{fig:magnetic-based jamming}
\end{figure}

% Overall design + membrain fabrification --> figure 2EA
The magnetic jamming module was designed to meet a set of qualitative goals related to user interaction and mechanical adaptability. First, the system should enable real-time stiffness switching to provide immediate haptic feedback to the user. Second, in the low-stiffness mode, the module should remain highly compliant, allowing it to flexibly conform to external deformations. Lastly, in high-stiffness mode, this enables the system to effectively lock into a user-defined shape, preserving the intended configuration.

To achieve these objectives, the module incorporates three core components: a flat-faced cylindrical electromagnet for magnetic field generation; a soft jamming core composed of mixed-size magnetic iron particles blended with silicone oil; and a deformable membrane that enables both surface conformity and magnetic responsiveness.

In fabrication, we used a cylindrical electromagnet ($\varnothing \ 50 \ mm$ DC 24 V) with a maximum holding force of 30 $kgf$ and a rated current of 0.26\,A. The jamming particles were composed of fine iron powder (particle size $ < 850 \ mm $ -20 mesh), iron granules ($1-2 \ mm$), and KF-96 silicone oil (ShinEtsu), blended at an optimized ratio. The membrane was first formed by preparing a mold with an internal geometry tailored to the desired shape and thickness. We made the mold using a 3D printer (Markforged X3) and plastic material (Onyx). A mixture of fine iron powder and silicone elastomer (Ecoflex 00-20) was poured into the mold to form the magneto-responsive membrane. The outer housing was constructed using laser-cut acrylic panels shaped to fit the system geometry, and the entire assembly was sealed with a hot-melt adhesive to ensure water and dust resistance (see Fig. \ref{fig:enter-label}).

%  Image
\begin{figure*}[t]
\center
\includegraphics[width=0.9\textwidth]{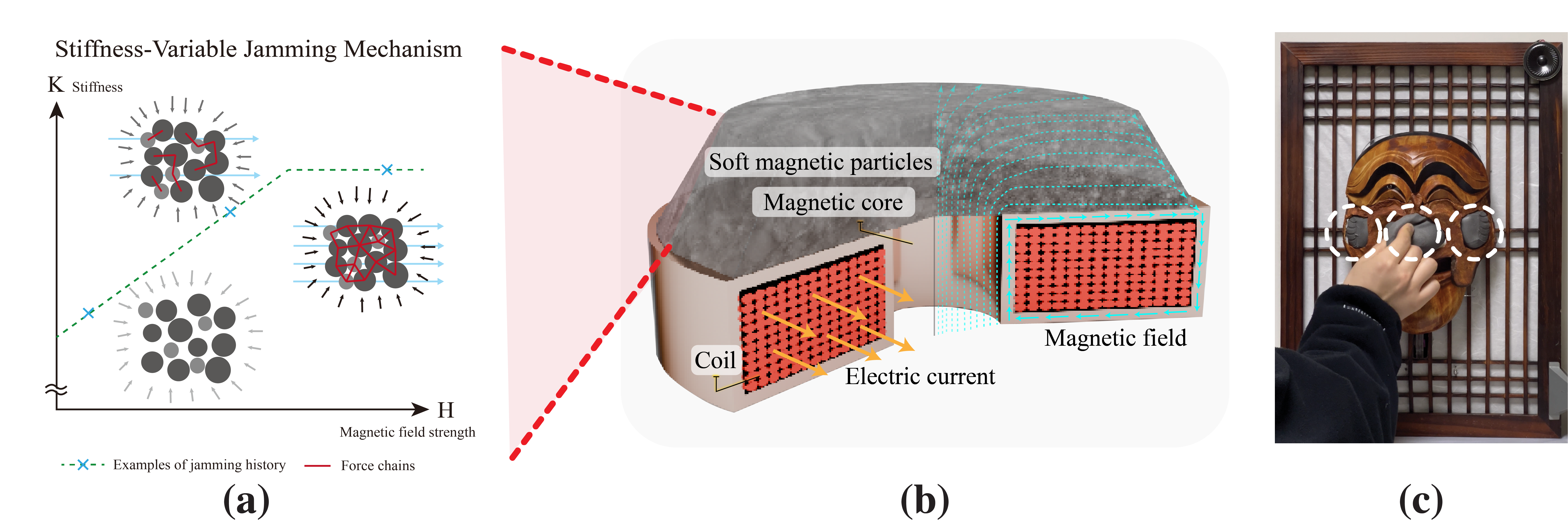}
\caption{
Our magnetic jamming mechanism is illustrated as follows:
\textbf{(a, b)} Electric current flowing through the magnetic core generates a magnetic field, which magnetizes the soft magnetic particles. As a result, the particles acquire attractive forces among themselves, forming a quasi-linear magnetic force. Despite the inherent anisotropy and weaker particle-particle interactions in the direction perpendicular to the field, our magnetic jamming module exhibits non-fragile stiffness due to its closed magnetic field design.
\textbf{(c)} Three magnetic jamming modules are applied to create artworks with varying rigidity, enabling dynamic tactile interactions and aesthetic experiences.
}
\label{img:method}
\end{figure*}

\subsection{Working Principle}
% Particles & membrain(force chain + ratio of particles) --> Figure force chain & Table of ratio and weight
The working principle of the magnetic jamming module relies on the synergistic interaction between iron particles and the surrounding membrane under a dynamically applied magnetic field as shown in Fig. \ref{img:method} (a) and (b). Upon activation, the electromagnet generates a closed-loop magnetic flux that permeates the iron particles dispersed within the module. This magnetic flux magnetizes the particles, inducing attractive forces that lead to the formation of quasi-linear magnetic force chains aligned along the direction of the magnetic field \cite{cates1998jamming}.

The strength of these force chains can be actively modulated by adjusting key parameters such as the magnetic field intensity and the size distribution of the iron particles. Stronger magnetic fields increase particle magnetization and inter-particle attraction, resulting in tunable stiffness and dynamic response of the system. And optimally mixed particle sizes enhance flux permeability and mechanical interlocking.

\subsection{Performance Characterization}
\begin{figure}[t]
\includegraphics[width=\columnwidth]{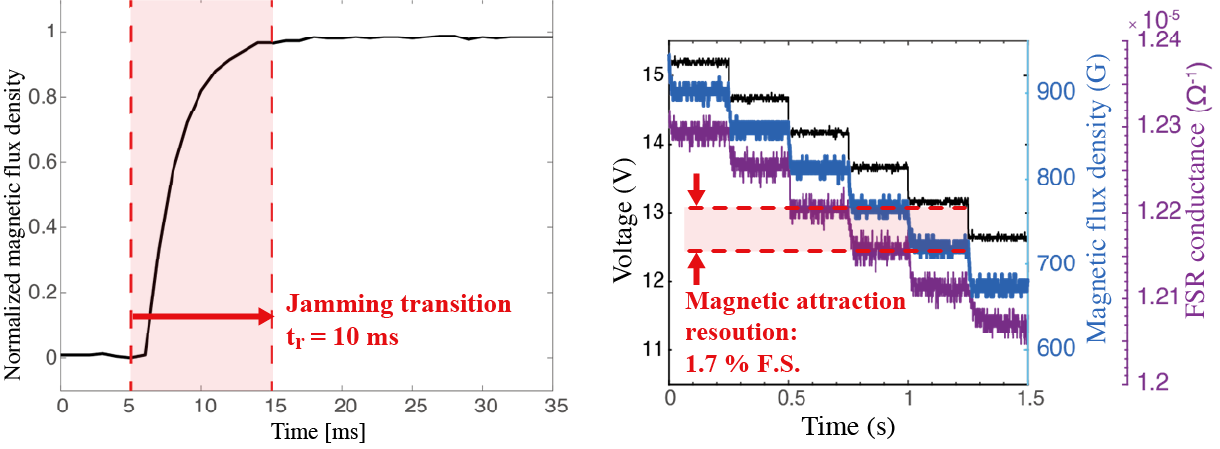}
\caption{
Characterizations of our jamming tangible module. \textbf{(Left)} Jamming transition time within 0.1 seconds.
\textbf{(Right)} Resolution of magnetic attraction response time with the step input of 0.5 V.
}
\label{img:graph}
\end{figure}

% \cy{figure 언급할 때 "as shown in Fig. \ref{}", "as detailed in Fig. \ref{}", (see Fig. \ref{}) 로 표현해주세요}

% chic and nose are different stiffness & elastic -> two independent properties of MBJM

% \usepackage{multirow}
% \usepackage{graphicx}
We fabricated two distinct modules – the “Nose” and the “Cheek” – to investigate the effects of parameter modifications on the mechanical performance and tunability of the magnetic jamming module (see Fig. \ref{fig:magnetic-based jamming}).

\begin{enumerate}
    \item \textbf{Magnetic Field Response Speed and Strength Resolution:} The magnetic flux density changes within 10 $ms$ as detailed in Fig. \ref{img:graph}, enabling users to experience immediate tactile feedback through rapid stiffness modulation. Additionally, the system allows fine control over a wide range of stiffness levels, providing versatile and precise strength adjustments during the jamming process.

    \item \textbf{Stiffness Variation Ratio (SVR) and Material Composition:} The SVR depends on the amount of oil and the iron particle size ratio, which affects the filling fraction\cite{jaeger1996granular}. More oil reduces friction and lowers stiffness. The highest SVR is achieved at 70\% packing, with increased powder content raising stiffness. The Nose module exhibits a higher SVR due to its greater oil content and a 7:3 granular-to-powder ratio, whereas the Cheek module, with less oil and a balanced 1:1 ratio, demonstrates a lower SVR but stronger locking force, as detailed in Table \ref{tab:test_table}.

    \item \textbf{Membrane Elasticity and Composition:} The membrane’s composition influences both the flexibility and magnetic responsiveness. For instance, the Nose module employs a higher proportion of Ecoflex, resulting in a membrane that is highly stretchable and flexible yet exhibits reduced magnetic sensitivity. Conversely, the cheek module’s membrane contains a higher concentration of iron powder, which increases stiffness and enhances shape fixation. Detailed composition weight in Table \ref{tab:test_table}.
\end{enumerate}

By systematically adjusting these parameters, each module can be customized to precisely meet specific performance requirements. However, due to physical limitations, the membrane requires external assistance, such as gravity or other forces, to fully restore its original shape after the jamming module is deactivated.

\begin{table}[]
\centering
\caption{material composition of jamming modules (unit: $g$)}
\renewcommand{\arraystretch}{1.2} 
\label{tab:test_table}
\resizebox{0.9\columnwidth}{!}{%
\begin{tabular}{cc|c|c|c|c}
\hline
\multicolumn{2}{c|}{} &
  \textbf{\begin{tabular}[c]{@{}c@{}}Iron\\ Granular\end{tabular}} &
  \textbf{\begin{tabular}[c]{@{}c@{}}Iron\\ Powder\end{tabular}} &
  \textbf{Oil} &
  \textbf{Silicone} \\ \hline
\multicolumn{1}{c|}{\multirow{2}{*}{\textbf{Nose}}}  & \textbf{Membrain}  & -     & 15   & -   & 25   \\ \cline{2-6} 
\multicolumn{1}{c|}{}                                & \textbf{Particles} & 182  & 78   & 20 & -     \\ \hline
\multicolumn{1}{c|}{\multirow{2}{*}{\textbf{Cheek}}} & \textbf{Membrain}  & -     & 38.3 & -   & 12.7 \\ \cline{2-6} 
\multicolumn{1}{c|}{}                                & \textbf{Particles} & 38.5 & 38.5 & 3  & -     \\ \hline
\end{tabular}%
}
\end{table}
\begin{figure*}[t]
\center
\includegraphics[width=\textwidth]{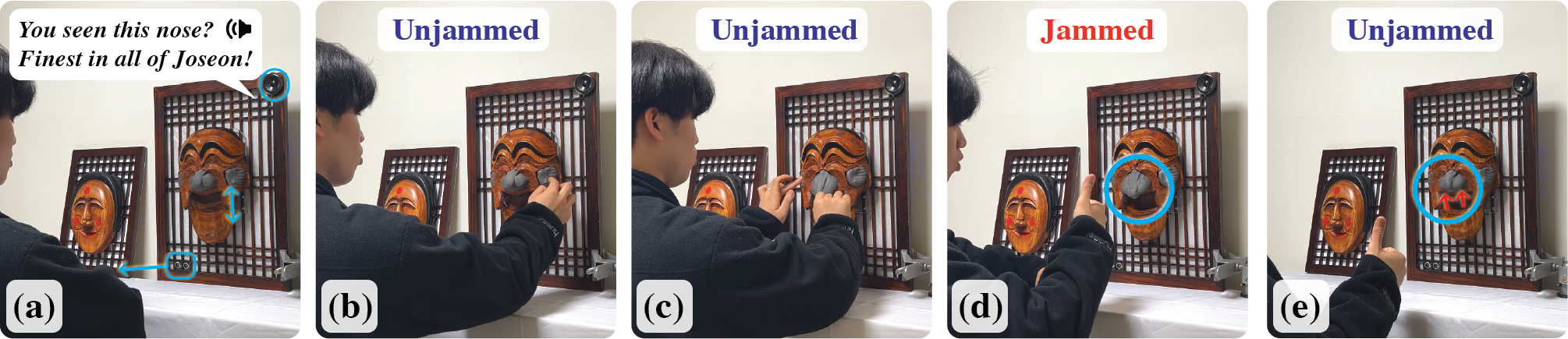}
\caption{
Illustration of the user interaction flow: (a) viewer approach detection followed by mask speech to attract users, (b) and (c) tactile exploration under low stiffness, (d) real-time stiffness increase, and (e) surface shape preservation and restoration.
}
\label{img:scinario}
\end{figure*}

\section{Interactive Artwork Application}
\subsection{Cultural Context: Hahoetal and Hanji}

 Korean traditional cultural heritage encompasses a variety of tangible and intangible assets, among which the \textit{Hahoetal} masks and \textit{Hanji} (traditional Korean paper) stand as prominent symbols of artistic and communal spirit. In this project, we propose the integration of these two cultural elements with interactive technology to create a novel form of multisensory cultural experience.
 
The \textit{Hahoetal} masks, originating from Hahoe Village in Andong, Gyeongsangbuk-do, are designated as National Treasure No. 121 and are part of the UNESCO World Heritage Site "Historic Villages of Korea: Hahoe and Yangdong" (2010) \cite{unesco_hahoe}. Historically used for both ritual and theatrical purposes, \textit{Hahoetal} masks vividly depict various social classes such as the \textit{Yangban} (aristocrats), \textit{Seonbi} (scholars), monks, and commoners through a unique blend of realistic and humorous artistry \cite{cha_hahoemasks}. One of their most notable features is the dynamic expressiveness of the masks; depending on the angle and movement of the wearer, the facial expressions appear to change fluidly, a rare trait among traditional sculptures. The associated \textit{Hahoe Byeolsingut Talnori} (Hahoe Mask Dance Drama), designated as Important Intangible Cultural Property No. 69 \cite{cha_hahoebyeolsingut}, was traditionally performed to pray for the well-being of the village and to satirize social hierarchies, thereby reflecting communal solidarity, social critique, and the spirit of resilience among Korean people.

\textit{Hanji}, made from the inner bark fibers of the mulberry tree, embodies centuries of traditional craftsmanship and communal collaboration. It has been utilized not only for record-keeping but also in architectural, artistic, and everyday applications \cite{jeong2015study}. Recognized for its exceptional durability and preservation qualities, \textit{Hanji} has contributed to the longevity of Korean documentary heritage, with works such as the \textit{Hunminjeongeum} and the Annals of the Joseon Dynasty being inscribed on UNESCO’s Memory of the World Register. Currently, \textit{Hanji} is undergoing the application process for inscription on UNESCO’s Representative List of the Intangible Cultural Heritage of Humanity.

By combining the dynamic expressiveness of \textit{Hahoetal} with the tactile richness of \textit{Hanji}, our project aims to transcend traditional passive viewing experiences. We enhance the interaction by introducing real-time deformation and stiffness modulation through a magnetic granular jamming mechanism, allowing the audience not only to see and hear but also to physically engage with the artwork. The system enables multisensory immersion where users can observe the mask's mouth motion, listen to auditory feedback, and touch surfaces that dynamically alter their rigidity. Furthermore, incorporating diverse textures such as wood, jamming-modulated surfaces, and \textit{Hanji} coverings into a single interactive artwork, we offer an enriched tactile experience that harmonizes Korea’s traditional aesthetic values with cutting-edge interactive technology.

\subsection{Concept: Fusion of Heritage and Modern Technology}
Traditional methods of appreciating cultural artifacts have primarily focused on passive visual observation. Although recent advances in digital technology and sensor-based systems have enabled more interactive exhibition formats, their application to Korean traditional arts, particularly sculpture-based heritage, remains limited. In particular, tactile engagement with traditional artworks has been largely static, offering surface-level interaction without dynamic changes or real-time responsiveness to user input. This project seeks to transcend these limitations by re-imagining traditional cultural experiences as living, interactive encounters. Specifically, the Korean traditional mask \textit{Hahoetal} is characterized by its dynamic expressiveness, with facial features that shift depending on the wearer’s movement and angle of view. Leveraging these intrinsic qualities, we aim to integrate \textit{Hahoetal} with modern interactive technologies to create a richer, emotion-driven engagement between the artwork and its audience.

 Our proposed system maintains the traditional appearance of the mask while enabling real-time emotional interaction with users. The system delivers a multisensory experience that combines real-time tactile feedback through stiffness and shape modulation, visual responses via mouth motion, and auditory feedback through voice interaction initiated by the mask. This integration of tactile, visual, and auditory modalities transforms the user experience from passive observation into active, immersive participation. By enabling viewers to see, touch, and hear the living dynamics of the mask, our approach offers a novel way to revitalize traditional culture, allowing heritage to breathe and respond anew through contemporary interactive technology.

\subsection{User Interaction Flow}
 The user interaction process with the proposed system is illustrated in the scenario shown in Fig. \ref{img:scinario}. 
 
 Initially, as the viewer approaches the artwork, an ultrasonic distance sensor detects their presence, triggering the \textit{Hahoetal} mask to respond. The mask animates its chin and initiates a voice interaction, delivering the following message: \textit{"Well now, you’ve seen my face—there’s no un-seein’ it, kid. You really think you can just stroll on by like nothin’ happened? Think again!"} (see Fig. \ref{img:scinario} (a)). 
 
 Following this interaction, the mask encourages the viewer to touch its nose, suggesting that it feels lifelike. At this stage, the jamming module remains deactivated, allowing the viewer to experience a soft and deformable surface when applying pressure to the nose and cheek regions (see Fig. \ref{img:scinario} (b)). 
 
 After a predefined time interval, the microcontroller reactivates the jamming modules by supplying voltage, leading to a rapid increase in surface stiffness. As a result, the deformed shapes of the nose and cheeks are preserved, enabling the viewer to directly experience the transition of the artwork’s stiffness and form in real time (see Fig. \ref{img:scinario} (b), (c), (d)). 
 
 Subsequently, when the jamming modules are deactivated again, the surface gradually returns to its original soft state (see Fig. \ref{img:scinario} (e)). Through this sequence, viewers experience the process of observing, deforming, and restoring the mask’s shape in real time, thereby experiencing an interactive exhibition that transcends passive observation and incorporates tactile interaction and dynamic material responses. This interaction paradigm redefines the relationship between viewers and traditional artworks. 

\section{Conclusion}

Integration of visual and tactile feedback is crucial for enhancing audience experiences in the arts. To achieve this, we propose a novel tangible display with variable stiffness, utilizing a magnetic jamming approach. Unlike previous tangible display approaches, our method enables real-time interaction with artworks without relying on bulky mechanical components, generating noise, or experiencing prolonged transition times.

As an application, we introduce a novel artwork that integrates a magnetic jamming-based 3D display with traditional korean masks, such as \textit{Hahoetal}, to provide audiences with an immersive and dynamic experience. Our magnetic jamming mechanism allows viewers to perceive variations in the rigidity of facial features, such as the Hahoetal's nose and cheeks, in real time, thereby enhancing their appreciation of the artwork.

\bibliographystyle{IEEEtran}

\IEEEtriggeratref{20} 

\bibliography{bib/reference}

\begin{IEEEbiography}[{\includegraphics[width=1in,height=1.25in,clip,keepaspectratio]{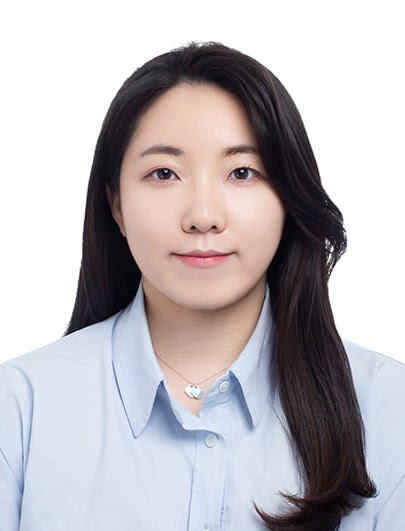}}]{Chanyoung Ahn} received the B.Eng. degree in Mechanical Engineering from Pusan National University, Busan, South Korea, in 2021 and the M.Eng. degree in robotics program from Korea Advanced Institute of Science and Technology (KAIST), Daejeon, South Korea, in 2024. She is currently working as a researcher at the Center for Humanoid Research, Korea Institute of Science and Technology (KIST), Seoul, South Korea.

Her research interests include dexterous manipulation and deformable object manipulation with reinforcement learning.
\end{IEEEbiography}
\vspace{-1.2em} % <<< 간격 줄이기 (필요에 따라 값 조정)

\begin{IEEEbiography}
[{\includegraphics[width=1in,height=1.25in,clip,keepaspectratio]{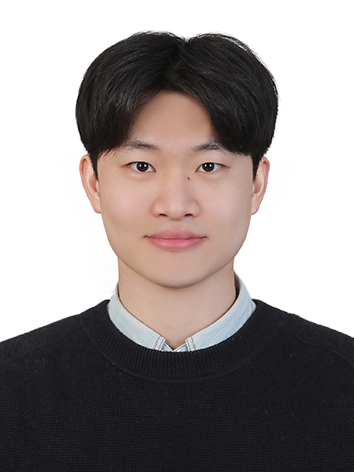}}]{Jaesung Lee} received the B.S. degree in Mechanical Engineering from Inha University, Incheon, South Korea, in 2025. He is currently working as a research intern at the Center for Humanoid Research, Korea Institute of Science and Technology (KIST), Seoul, South Korea.

His research interests include pressure mapping sensors, robotic grippers, and intelligent hand control.
\end{IEEEbiography}

\vspace{-1.2em} % <<< 간격 줄이기 (필요에 따라 값 조정)

\begin{IEEEbiography}
[{\includegraphics[width=1in,height=1.25in,clip,keepaspectratio]{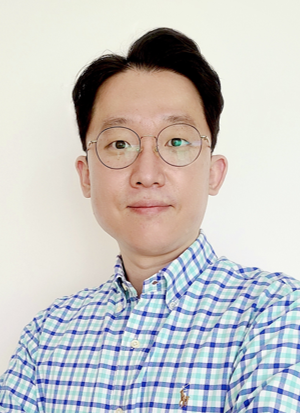}}]{Donghyun Hwang (Member, IEEE)}
received the B.Eng. and M.Sc. degrees in Mechanical Engineering from Ajou University, Suwon, South Korea, in 2008 and 2010, respectively, and the Ph.D. degree in precision engineering from The University of Tokyo, Tokyo, Japan, in 2014. In 2010, he was a Visiting Researcher with the Department of Mechanical Engineering, University of Michigan, Ann Arbor, MI, USA. From 2014 to 2015, he was a Postdoctoral Research Fellow with the Center for Robotics Research, Korea Institute of Science and Technology, Seoul, South Korea, where he is currently a Principal Research Scientist. 

His research interests include smart actuators and sensors, variable stiffness compliant mechanisms, ultra-precision positioning systems for robotic applications including microsurgical robotic systems, haptic interfaces, wearable devices, and robotic manipulators.
\end{IEEEbiography}

\end{document}